%% file: masterfile.tex
\documentclass[10pt,twoside,BCOR7mm,DIV15,headinclude,footexclude,cleardoubleempty,idxtotoc]{scrartcl}

\usepackage[english]{babel}
\usepackage{graphicx}
\usepackage{hyperref}
\usepackage{scrpage2}
\usepackage{hyperref}
\usepackage{ifthen}

\makeatletter
\renewcommand{\@biblabel}[1]{}
\renewcommand{\@cite}[2]{%
{#1\ifthenelse{\boolean{@tempswa}}{,#2}{}}}
\makeatother

\hypersetup{breaklinks=true
,colorlinks=true,linkcolor=black,urlcolor=blue
,citecolor=black}

\ofoot{\thepage}
\ifoot{}

\setheadsepline{1pt}

\setkomafont{pagehead}{\normalfont\sffamily}
\setkomafont{pagenumber}{\normalfont\rmfamily}

\usepackage{booktabs}
\usepackage{amsmath}
\usepackage{amssymb}
\usepackage{multicol}
\usepackage{float}

\makeatletter
\newcommand{\listofcontributions}{\@starttoc{con}}

\newcommand{\l@contribution} {\@dottedtocline{1}{1.5em}{2.3em}}
\makeatother

\newenvironment{contribution}{
\setcounter{section}{0}
\setcounter{figure}{0}
\setcounter{table}{0}
\begin{flushleft}
{\em Clumping in Hot Star Winds \\
W.-R.\ Hamann, A.\ Feldmeier \& L.\ Oskinova, eds.\\
Potsdam: Univ.-Verl., 2007 \\
URN: http://nbn-resolving.de/urn:nbn:de:kobv:517-opus-13981
} 
\end{flushleft}
}{
\newpage
\lehead{}
\rohead{}
}

\begin{document}

\setlength{\baselineskip}{2.5ex}

\begin{contribution}
\input{myarticle.tex}
\end{contribution}


\end{document}

%% file: myarticle.tex
\lehead{R. Walter, J. Zurita-Heras \& J.-C. Leyder}

\rohead{Probing clumpy stellar winds with a neutron star}

\begin{center}
{\LARGE \bf Probing clumpy stellar winds with a neutron star}\\
\medskip

{\it\bf R.\ Walter$^{1,2}$, J.\ Zurita-Heras$^3$ \& J.-C.\ Leyder$^4$}\\

{\it $^1$Observatoire de Gen\`eve, Universit\'e de Gen\`eve, Chemin des Maillettes 51, CH-1290 Sauverny}\\
{\it $^2$INTEGRAL Science Data Centre, Chemin d'Ecogia 16, CH-1290 Versoix}\\
{\it $^3$Laboratoire AIM, CEA/DSM-CNRS-Universit\'e Paris Diderot, DAPNIA/SAP, F-91191 Gif-sur-Yvette}\\
{\it $^4$Institut d'Astrophysique et de G\'eophysique, Universit\'e de Li\`ege, All\'ee du 6-Ao\^ut 17, B-4000 Li\`ege}

\begin{abstract}
INTEGRAL tripled the number of super-giant high-mass X-ray binaries (sgHMXB) known in the Galaxy by revealing absorbed and fast transient  (SFXT) systems.
Quantitative constraints on the wind clumping of massive stars can be obtained from the study of the hard X-ray variability of SFXT.
A large fraction of the hard X-ray emission is emitted in the form of flares with a typical duration of 3 ksec, frequency of 7 days and luminosity of $10^{36}$ ergs/s. Such flares are most probably emitted by the interaction of a compact object orbiting at $\sim10~R_*$ with wind clumps ($10^{22-23}$ g) representing a large fraction of the stellar mass-loss rate. The density ratio between the clumps and the inter-clump medium is  $10^{2-4}$. 
The parameters of the clumps and of the inter-clump medium, derived from the SFXT flaring behavior, are in good agreement with macro-clumping scenario and line-driven instability simulations. SFXT are likely to have larger orbital radius than classical sgHMXB.
\end{abstract}
\end{center}

\begin{multicols}{2}

\section{Introduction}

Indirect measures of the structure of massive-star winds are possible in X-ray binaries through the analysis of the interaction between the compact companion and the stellar wind. 
In this  report we summarize the constraints obtained on wind clumping in HMXB using the hard X-ray variability observed by the IBIS/ISGRI instrument 
on board INTEGRAL (\cite{walter:winkler03AA}). Further details can be found in \cite{walter:WalterZurita2007} and \cite{walter:Leyder2007}.

Classical wind-fed, Roche-lobe underflow, super-giant HMXB (sgHMXB) are made of a compact object orbiting within a few (1.5 to 2.5) stellar radii from a super-giant companion. Recently INTEGRAL almost tripled the number of sgHMXB systems known in the Galaxy and revealed a much more complex picture with two additional families of sources:
(1) the highly-absorbed systems which have orbital and spin periods similar to those of classical  sgHMXB but much higher absorbing column densities on average (\cite{walter:walter2006}) and (2) the fast transient systems which are characterized by fast outbursts and by a very low quiescent luminosity (\cite{walter:Sguera2006,walter:Negueruela2007}).

\section{Sources and data analysis}

Several sources have now been proposed as candidate super-giant fast X-ray transient based on their hard X-ray variability characteristics, and, for a subset of them, optical counterpart spectral type. Contrasting statements have however been made on specific sources for what concerns their persistent or transient nature. In the frame of the current study we have considered all SFXT candidates together with several persistent and absorbed super-giant HMXB for comparison. 
Among them, we specifically excluded known Be systems, sources detected only once by INTEGRAL,
blended INTEGRAL sources, long period systems and the sgB[e] system IGR J16318$-$4848.
 
We analyzed the available INTEGRAL data for 12 candidate SFXT (table \ref{walter:tab2}) that have large variability factors and compared them with the classical and absorbed sgHMXB systems that have a typical variability factor $\lesssim20$.
The sources of the sample are located along the galactic plane that has been heavily observed by INTEGRAL. All public data available until March 2007 are considered in this study. 
Individual ISGRI sky images have been produced for each INTEGRAL pointing in the energy band 22--50~keV. The detection of the sources of the sample is forced in each image and the source count rate extracted. 

Source flares have been detected by requiring a minimum of 25 ksec of inactivity between them. Flare duration of the order of a single INTEGRAL pointing $(2~\rm{ksec})$ have been observed in all sources (excepting IGR\,J16465$-$4507). Their typical duration is 3 ksec. 
Fewer longer $(> 15~\rm{ksec})$ flares have also been detected but in most cases could be interpreted as a serie of shorter flares or a long activity period. They will not be discussed further here. 

\begin{table}[H]
\vspace{-5mm}

\caption{List of SFXT candidates with quiescent flux $F_{q}$, source observing elapsed time $T_{obs}$ and flaring characteristics: 
maximum count rate $F_{fl}$, number of flares $N_{fl}$ and the average flare duration $t_{fl}$.} 
\label{walter:tab2}
\begin{tabular}{l|r@{.}l|c|c|l|r}
\noalign{\vspace{2mm}}
\toprule 
Source                   &\multicolumn{2}{c|}{$F_{q}$}&\multicolumn{1}{c|}{$F_{fl}$}&$N_{fl}$&\multicolumn{1}{c|}{$t_{fl}$}&\multicolumn{1}{c}{$T_{obs}$}\\
                  &\multicolumn{2}{c|}{ct/s}&\multicolumn{1}{c|}{ct/s}&&\multicolumn{1}{c|}{ks}&\multicolumn{1}{c}{days}\\
\midrule 
\noalign{\vspace{2mm} \bf SFXT systems\vspace{2mm}}
\tiny{IGR\,J08408$-$4503}     &$<0$&1&3.9  &2    &3.6               &52.0\\
\tiny{IGR\,J17544$-$2619}     &     0&06   &24&8&2.5   &127.0\\
\tiny{XTE\,J1739$-$302}         &     0&08   &28&12 &4.2   &126.4\\
\tiny{SAX\,J1818.6$-$1703}   &     0&18   &45&11    &2.9   &76.9\\
\tiny{IGR\,J16479$-$4514}     &     0&2   &19&38 &3.6  &67.0\\
\tiny{AX\,J1841.0$-$0536}      &$<0$&1&15&4     &5.8&51.9\\
\tiny{AX\,J1820.5$-$1434}      &$<0$&1&5.3&4     &3.9 &59.4\\
\noalign{\vspace{2mm} \bf Intermediate systems\vspace{2mm}}
\tiny{AX\,J1845.0$-$0433}       &    0&2  &6.2 &6    &4.0&55.2\\
\tiny{IGR\,J16195$-$4945}      &    0&2  &4.8 &6    &2.2&71.8\\
\tiny{IGR\,J16465$-$4507}      &     0&1  &6.9&3    &     &66.7\\
\tiny{IGR\,J16207$-$5129}      &    0&4  &9.2 &11  &4.3&73.7\\
\tiny{XTE\,J1743$-$363}          &     0&5  &9.2 &19&2.5&122.9\\
\bottomrule
\end{tabular}
\end{table}

Table \ref{walter:tab2} lists the sources together with their quiescent count rate $(F_{q})$, average flare count rate $(F_{fl})$, number of flares $(N_{fl})$, range of flare durations $(t_{fl})$ and total source observing elapsed time $(T_{obs})$. 
As the probability to detect a flare decreases when the source gets outside of the fully-coded field of view, the effective observing time for flare detection can be estimated as $0.6~T_{obs}$. 

The sources have been separated in two categories. The SFXT include systems featuring hard X-ray variability by a factor $\gtrsim100$. ``Intermediate'' systems are candidate SFXT with smaller variability factors that could be compared with those of classical systems. 
From the variability point of view, sources closer to the bottom of the table are more similar to classical sgHMXB.

\section{Discussion}

The distances to the SFXT systems has been evaluated (2--7 kpc) in a few cases. 
We will assume, for the rest of the discussion, a distance of 3 kpc. The average count rate observed during flares lies between 3 and 60 ct/s which translates to hard X-ray luminosities of $(0.2-4)\times 10^{36}~\rm{erg/s}$. Such luminosities are not exceptional for sgHMXB but very significantly larger than the typical X-ray luminosity of single massive stars of $10^{30-33}~\rm{erg/s}$ at soft X-rays (\cite{walter:Cassinelli1981}). 

As the sources are flaring at most once per day, their average hard X-ray luminosity is very low, reaching $(0.2-4)\times 10^{34}~\rm{erg/s} $. It is therefore very unlikely that those systems have average orbital radius lower than $10^{13}~\rm{cm}$ i.e. $\sim 10~R_*$. One expects orbital periods larger than 15 days and underflow Roche lobe systems (note that no orbital period has yet been derived in any of these systems). 

\paragraph {Wind clumps}
\paragraph {}
\vspace{-0.3cm}
The interaction of a compact objet with a dense clump formed in the wind of a massive companion leads to increased accretion rate and hard X-ray emission. 

The free-fall time from the accretion radius $R_a = 2\times 10^{10}~ \rm{cm}$ towards the compact object is of the order of $(2-3)\times10^2~\rm{sec}$. As the intrinsic angular momentum of the accreted gas is small (\cite{walter:Illarionov2001}) the infall is mostly radial (down to the Compton radius) and proceeds at the Bondi-Hoyle accretion rate. 

With a duration of $t_{fl}=2-10$ ksec, the observed short hard X-ray flares are significantly longer than the free-fall time. The flare duration is therefore very probably linked with the thickness of the clumps which,  for a clump radial velocity $V_{cl}=10^8 ~\rm{cm/s}$, is $h_{cl} = V_{cl} \times t_{fl} \sim (2-10) \times 10^{11}~\rm{cm}$.

The average hard X-ray luminosity resulting from an interaction between the compact object and the clump can be evaluated as $L_X  = \epsilon~M_{acc}c^2/t_{fl}$ (where $\epsilon\sim0.1$) and the mass of a clump can then be estimated as
$ M_{cl} = ~ (R_{cl}/R_{a})^2 ~M_{acc}= (R_{cl}/R_{a})^2~L_X~t_{fl}/(\epsilon~ c^2) $
where $R_{cl}$ is the radius of the clump perpendicular to the radial distance.
In the case of a spherical clump,
$M_{cl} = 
\left(\frac{L_X}{10^{36}~\rm{erg/s}}\right) \left(\frac{t_{fl}}{3~\rm{ks}}\right)^3 
~7.5\times 10^{21} ~\rm{g}.$

If $\dot{N}$ is the rate of clumps emitted by the star, the observed hard X-ray flare rate is given by $T^{-1} = \dot{N}(R_{cl}^2/4R_{orb}^2).$
The rate of mass-loss in the form of wind clumps can then be estimated as
$\dot{M}_{cl}  =
\left(\frac{10\rm{d}}{T}\frac{L_X}{10^{36}\rm{erg/s}}\frac{t_{fl}}{3\rm{ks}}\right)\left(\frac{R_{orb}}{10^{13}\rm{cm}}\right)^2 ~3\times 10^{-6}~\rm{M_{\odot}/y}.$

For a $\beta=1$ velocity law and spherical clumps, the number of clumps located between $1.05R_*$ and $R_{orb}$  can be evaluated as  
$N=
\left(\frac{10~\rm{d}}{T}\right)\left(\frac{3~\rm{ks}}{t_{fl}}\right)^2\left( \frac{R_{orb}}{10^{13}~\rm{cm}}\right)^3~3.8\times 10^3$. 

Assuming spherical clumps, the clump density at the orbital radius is $\rho_{cl}=\left(\frac{L_X}{10^{36}~\rm{erg/s}}\right) ~7\times 10^{-14} ~\rm{g~cm}^{-3}$ and the corresponding homogeneous wind density is $\rho_h=\dot{M}_{cl}/(4\pi~R_{orb}^2~V_{cl})=
\left(\frac{10~\rm{d}}{T}\frac{L_X}{10^{36}~\rm{erg/s}}\frac{t_{fl}}{3~\rm{ks}}\right)
~1.5\times 10^{-15}~\rm{g~cm}^{-3}$. The clump volume filling factor at the orbital radius is $
f_V = \frac{\rho_h}{\rho_{cl}} = 
\left(\frac{10~\rm{d}}{T}\frac{t_{fl}}{3~\rm{ks}}\right)
~0.02$ and the corresponding porosity length is
$h=\frac{R_{cl}}{f_V}=
\left(\frac{T}{10~\rm{d}}\right)
~15\times 10^{12} ~\rm{cm}$.

If the density of a clump decreases with radius as $r^{-2\beta}$ and its mass remains constant, the averaged homogeneous wind density within $R_{obs}$ is  $\overline{\rho_{h}}=N M_{cl}/(\frac{4}{3}\pi 
R_{orb}^3
) = 
\left(\frac{10~\rm{d}}{T}\frac{L_X}{10^{36}~\rm{erg/s}}\frac{t_{fl}}{3~\rm{ks}}\right)
~7\times 10^{-15} ~\rm{g~cm}^{-3}$ and the average clump volume filling factor and porosity length could be estimated as 0.1 and $3\times10^{12} ~\rm{cm}$, respectively.

The variety of $t_{fl}$, $T$ and $F_{fl}$ that are observed probably reflects a range of  clump parameters and orbital radii. Several of the average clump parameters estimated above, in particular the clump density, filling factor and  porosity length do not depend on the orbital radius, which is unknown, and only slowly depend on the observed quantities.
 
These average parameters match the macro-clumping scenario of \cite{walter:OskinovaHamannFeldmeier2007}
to reconcile clumping and mass-loss rates. 
The number of clumps derived above is also comparable to evaluations by \cite{walter:Lepine1999, walter:OskinovaFeldmeierHamann2006}. The volume filling factor, porosity length and the clump mass-loss rate are also similar to those derived by \cite{walter:Bouret2005} from the study of ultraviolet and optical line profiles in two super-giant stars.

The column density through a clump can also be estimated as $N_H = \frac{M_{cl}}{R_{cl}^2m_p}=
\left(\frac{L_X}{10^{36}\rm{erg/s}}\frac{t_{fl}}{3\rm{ks}}\right)
~5\times 10^{22}\rm{cm}^{-2}$. The clumps remain optically thin in the X-rays.

\paragraph{Inter-clump medium}
\paragraph{}
\vspace{-0.3cm}
The variation of the observed X-ray flux between flares and quiescence provides in principle a direct measure of the density constrast between the wind clumps and the inter-clump medium. 

Density contrasts of $>10^{2-4}$ and 15--50 have been observed in SFXT and ``Intermediate'' sources, respectively. The density contrast is larger in SFXT than in ``Intermediate'' and, of course, classical systems. Density contrasts are probably stronger when clumping is very effective. 

Numerical simulations of the line driven instability (\cite{walter:Runacres2005}) predict density contrasts as large as $10^{3-5}$ in the wind up to large radii. At a distance of $10~R_*$, the simulated density can vary between $10^{-18}$ and $10^{-13}~\rm{g~cm^{-3}}$ and the separation between the density peaks are of the order of  $R_*$. These characteristics are comparable to the values we have derived.

\paragraph{What about classical sgHMXB ?}
\paragraph{}
\vspace{-0.3cm}
Classical sgHMXB are characterized by small orbital radii $R_{orb}=(1.5-2.5)~R_*$, and by flux variability of a factor $\lesssim10$. Such variabilities were modelled in terms of wind inhomogeneities largely triggered by the hydrodynamic and photo-ionisation effects of the accreting object on the companion and inner stellar wind (\cite{walter:blondin91, walter:blondin94}). At small orbital radii, the companion is close to fill its Roche lobe, which triggers tidal streams. In addition the X-ray source ionizes the wind acceleration zone, prevents wind acceleration and generates slower velocities, denser winds, larger accretion radius and finally larger X-ray luminosities. Whether or not the stellar wind is intrinsically clumpy at low radius, the effect of the compact object on the wind is expected to be important.

The main difference between SFXT and classical sgHMXB could therefore be their orbital radius  (\cite{walter:Leyder2007}). At very low orbital radius $(<1.5~R_*)$ tidal accretion will take place through an accretion disk and the system will soon evolve to a common envelope stage. At low orbital radius $(\sim 2~R_*)$ the wind will be perturbed in any case and efficient wind accretion will lead to copious and persistent X-ray emission $(10^{36-37}~\rm{erg/s})$. At larger orbital radius $(\sim 10~R_*)$ and if the wind is clumpy, the SFXT behavior is expected as described above. If the wind clumps do not form for any reason, the average accretion rate will remain too low and the sources will remain mostly undetected by the current hard X-ray survey instruments.

\end{multicols}